\documentclass{aastex631}
\usepackage[utf8]{inputenc}
\usepackage{CJK}
\usepackage{amsmath}

\usepackage{booktabs}
\usepackage{tablefootnote}
\usepackage{multirow}

\newcommand{\mum}{\ensuremath{\,{\rm \mu m}}}

\newcommand{\K}{\ensuremath{\,{\rm K}}}

\newcommand{\s}{\ensuremath{\,{\rm s}}}
\newcommand{\g}{\ensuremath{\,{\rm g}}}

\shorttitle{IR Extinction in M16}
\shortauthors{Li et al.}

\begin{document}
	\begin{CJK*}{UTF8}{gbsn}
		
		\title{Probing the distinct extinction law of the Pillars of Creation in M16 with JWST}
		
		\author[0000-0001-9328-4302]{Jun Li (李军)}
		\affiliation{Center for Astrophysics, Guangzhou University, Guangzhou 510006, People's Republic of China}
		
		\author[0000-0003-2472-4903]{Bingqiu Chen (陈丙秋)}
		\affiliation{South-Western Institute for Astronomy Research, Yunnan University, Kunming, Yunnan 650091, Peopleʼs Republic of China}

		\author[0000-0003-3168-2617]{Biwei Jiang (姜碧沩)}
		\affiliation{Institute for Frontiers in Astronomy and Astrophysics, Beijing Normal University, Beijing 102206,  People's Republic of China}
		\affiliation{Department of Astronomy, Beijing Normal University, Beijing 100875, People's Republic of China}
		
		\author[0000-0003-4195-0195]{Jian Gao (高健)}
		\affiliation{Institute for Frontiers in Astronomy and Astrophysics, Beijing Normal University, Beijing 102206,  People's Republic of China}
		\affiliation{Department of Astronomy, Beijing Normal University, Beijing 100875, People's Republic of China}

		\author[0000-0002-5435-925X]{Xi Chen (陈曦)}
		\affiliation{Center for Astrophysics, Guangzhou University, Guangzhou 510006, People's Republic of China}
		
		\correspondingauthor{Bingqiu Chen, Xi Chen}
		\email{bchen@ynu.edu.cn, chenxi@gzhu.edu.cn}

		\begin{abstract}
			Investigating the extinction law in regions of high dust extinction, such as the Pillars of Creation within the M16 region, is crucial for understanding the densest parts of the interstellar medium (ISM). In this study, we utilize observations from the Near-Infrared Camera (NIRCam) and the Mid-Infrared Instrument (MIRI) onboard the James Webb Space Telescope (JWST) to analyze the color-excess ratios $E(F090W-\lambda)/E(F090W-F200W)$ across a wavelength range of $0.9-7.7\,\mu\mathrm{m}$. Our method involves performing linear regression on color-color diagrams to derive these ratios. The enhanced detection capabilities of JWST data allow us to probe the distinct extinction law to the densest regions in M16 corresponding to an extinction depth up to $A_V \sim 60$\,mag. Remarkably, the resultant color-excess ratio curve exhibits a flatter profile than predicted by typical dust extinction models with $R_V = 5.5$ for dense ISM environments. Moreover, we observe that the mid-infrared (MIR) extinction law diverges from the near-infrared (NIR) power-law, showing a tendency for the slope to flatten as the wavelength increases. These findings have significant implications for our understanding of the dust properties in dense interstellar environments.
		\end{abstract}
		
		
		\section{Introduction} \label{sec:intro}
		
		The extinction law characterizes the wavelength-dependent absorption and scattering of light by interstellar dust grains. This critical astrophysical parameter reveals the composition, size distribution, and structural properties of dust within the interstellar medium (ISM) \citep{Draine2003}. Dense regions within molecular clouds, where significant reddening occurs, are pivotal areas of study as they are the cradles of star and planet formation. The characteristics of the dust in these regions significantly impact the initial conditions for star formation and the evolution of protoplanetary disks. Thus, understanding the extinction law in these environments, although challenging, is essential.
		
		Extinction laws have been extensively studied in regions of the ISM with lower density, where extinction effects are less pronounced. The Galactic extinction curve at ultraviolet (UV) and optical wavelengths, which is often characterized in terms of the ratio $R_V = A_V/E(B-V)$, has revealed non-universal dust properties that differ across various sightlines \citep{Cardelli1989, Fitzpatrick2019, Massa2020}. In contrast, the near-infrared (NIR) extinction law, particularly from 1.2 to 2.2 \(\mu\)m, was previously considered uniform across the Galaxy, typically modeled by a power-law $A_\lambda \propto \lambda^{-\alpha}$ with $\alpha$ values around 1.6--1.8 \citep{Rieke1985,Cardelli1989}. However, more recent studies have reported $\alpha$ values ranging from 2 to 2.6 \citep{Stead2009, Alonso2017,Maiz2020}, and suggested a broader applicability of this power-law form across a wider wavelength range (0.8--5.0\mum) \citep{Decleir2022, Wang2024}. Furthermore, variations in the NIR power-law index both with wavelength and across different sightlines have been observed, challenging the previous assumption of a uniform NIR extinction law \citep{Alonso2017, Nogueras2019}. In the MIR range (3--8\mum), earlier studies posited that the NIR power-law could extend into the MIR wavelengths \citep{Rieke1985, Martin1990}. However, later observations, especially those from the Spitzer Space Telescope, have depicted a mostly flat extinction curve across these wavelengths in various interstellar environments \citep{Indebetouw2005, Gao2009, Chen2013, Xue2016}. This suggests that the MIR extinction law might significantly differ from the NIR laws, although observational challenges have historically limited a comprehensive understanding in this regime.
		
		Notably, the extinction depths investigated in most previous studies are not particularly high. However, in dense interstellar regions, variations in the extinction law are expected due to factors such as dust grain growth and ice-mantle formation, as suggested by both theoretical models \citep{Ossenkopf1994, Ormel2011} and observations \citep{Juvela2015,Li2024}. Observational studies indicate that in regions of significant star formation, where extinction is higher, the extinction law tends to flatten in comparison to more diffuse regions \citep{Moore2005, Flaherty2007, McClure2009}. This flattening aligns well with the \citeauthor{Weingartner2001} (2001, hereafter WD01) extinction model, which posits an $R_V=5.5$ for dense ISM environments. Additionally, there is accumulating evidence that suggests a trend toward a flatter IR extinction law with increasing extinction depth \citep{Chapman2009, Cambresy2011, Li2023}. 
		
		Characterizing the IR extinction law in dense environments presents significant challenges due to the high column densities and deep extinction that often obscure background stars. However, the advent of the James Webb Space Telescope (JWST) with its advanced instruments, Near Infrared Camera (NIRCam; \citealp{Rieke2005, Rieke2023}) and  Mid-Infrared Instrument (MIRI; \citealp{Rieke2015}), offers new opportunities. These instruments provide unprecedented sensitivity and spatial resolution, enabling a detailed examination of the extinction law across a broad range of IR wavelengths. In this paper, we focus on the IR extinction law in the Pillars of Creation within the Eagle Nebula (M16), a well-known star-forming region located at a distance of $1.74 \pm 0.13$\,kpc \citep{Kuhn2019}.
		
		The structure of this paper is organized as follows: Section \ref{sec:obser} details the observations and the methodology of data reduction. Section \ref{sec:results} discusses the results, focusing on the measurement and implications of the IR extinction law in M16. Section \ref{sec:summary} provides a summary of the findings and conclusions.
		
		\section{Observations and Data Reduction} \label{sec:obser}
		
		We utilize publicly available observations of M16 captured using both NIRCam and MIRI on board the JWST, as part of the Director's Discretionary (DD) program (ID: 2739; PI: Klaus Pontoppidan). The observations employed the NIRCam filters $F090W$, $F187N$, $F200W$, $F335M$, and $F444W$, along with the MIRI filters $F770W$, $F1130W$, and $F1500W$. These observations spanned the main structure of the Pillars of Creation in M16. The fields-of-view (FoVs) are 7.4$\times$4.4 arcmin$^2$ for NIRCam and 4.3$\times$3.8 arcmin$^2$ for MIRI, respectively.
		
		We have retrieved the Level-2b data products from the Mikulski Archive for Space Telescopes (MAST) \footnote{\protect\url{https://mast.stsci.edu}} at the Space Telescope Science Institute for data reduction. The specific data analyzed can be accessed via the MAST \dataset[DOI 10.17909/3w1e-qp71]{http://dx.doi.org/10.17909/3w1e-qp71}. The NIRCam images were processed using JWST pipeline version 1.10.1 with Calibration Reference Data System (CRDS) version 11.16.22 and context \texttt{jwst\_1100.pmap}. MIRI images were processed using JWST pipeline version 1.13.3 with CRDS version 11.17.14 and context \texttt{jwst\_1216.pmap}. Initially, the Level-2b \texttt{*i2d.fits} files were corrected for $1/f$ noise using the Python routine \texttt{image1overf.py}\footnote{\protect\url{https://github.com/chriswillott/jwst}} \citep{Willott2022}. Subsequently, we use the $F200W$ Level-3 drizzled \texttt{i2d.fits} file as the astrometric reference to extract point sources. The astrometric alignment is applied to all Level-2b \texttt{*i2d.fits} files using the JWST/Hubble Alignment Tool (JHAT\footnote{\protect\url{https://github.com/arminrest/jhat}}; \citealp{Rest2023}).
		
		For photometric analysis, we have performed PSF photometry on the NIRCam and MIRI data using the \textsc{starbugII} photometric tool \citep{Nally2023}, which is optimized for JWST observations in crowded fields with complex backgrounds. Source detection on the aligned Level-2b images is executed using \texttt{starbug2 --detect}, with a detection threshold of 5$\sigma$ for the NIRCam $F090W$, $F187N$, and $F200W$ sources, and 3$\sigma$ for the NIRCam $F335M$, $F444W$, and MIRI sources. We have applied criteria related to geometric parameters such as sharpness and roundness to filter out contaminants including cosmic rays, dust structures, and galaxies. The parameters used for \textsc{starbugII} detection and photometry are summarized in Table~\ref{tab:starbug_para}.
		
		\begin{table*}
			\centering
			\caption{\textsc{StarbugII} parameters used for photometry}
			\label{tab:starbug_para}
			\begin{tabular}{c c c c c c  c c c}
				\hline \hline
				Parameter &  $F090W$& $F187$N & $F200W$ & $F335M$ & $F444W$ & $F770W$ & $F1130W$ & $F1500W$ \\
				\hline
				SIGSKY & 2 & 2 & 2 & 1.5 & 1.5 & 1.4 & 1.2 & 1.2 \\
				SIGSRC & 5 & 5 & 5 & 3.0 & 3.0 & 3.0 & 3.0 & 3.0 \\
				RICKER\_R & 1 & 1 & 1 & 1 & 1 & 1 & 4 & 5 \\
				SHARP\_LO & 0.4 & 0.4 & 0.4 & 0.4 & 0.4 & 0.3 & 0.25 & 0.2 \\
				SHARP\_HI & 1.1 & 0.9 & 0.9 & 0.9 & 0.9 & 0.9 & 0.8 & 0.8 \\
				ROUND\_LO/HI & $\pm$1.5 & $\pm$1 & $\pm$1 & $\pm$1 & $\pm$1& $\pm$1 & $\pm$0.8 & $\pm$0.8 \\
				\hline
				APPHOT\_R & 1.5 & 1.5 & 1.5 & 1.5 & 1.5 & 2.5 & 2.5 & 3 \\
				SKY\_RIN & 3 & 3 & 3 & 3 & 3 & 4 & 4 & 4.5 \\
				SKY\_ROUT & 4.5 & 4.5 & 4.5 & 4.5 & 4.5 & 5.5 & 5.5 & 6 \\
				BOX\_SIZE & 15 & 15 & 15 & 5 & 5 & 5 & 5 & 5 \\
				CRIT\_SEP & 5 & 6 & 6 & 8 & 8 & 8 & 8 & 8 \\
				\hline
				MATCH\_THRESH & 0.06 & 0.06 & 0.06 & 0.1 & 0.1 & 0.15 & 0.2 & 0.2 \\
				NEXP\_THRESH & 2 & 2 & 2 & 2 & 2 & 3 & 3 & 3 \\
				\hline
			\end{tabular}
		\end{table*}

		Aperture photometry is also performed on all NIRCam and MIRI images to facilitate source detection and zero-point correction. For NIRCam, we have employed an aperture radius of 1.5 pixels with a sky annulus extending from 3.0 to 4.5 pixels. In contrast, for MIRI, aperture radii are set at 2.5 pixels for the F770W and F1130W bands and 3 pixels for the F1500W band, with background annuli ranging from 4 to 5.5 pixels and 4.5 to 6 pixels respectively. Aperture corrections are derived from \texttt{jwst\_nircam\_apcorr\_0004.fits} for NIRCam and \texttt{jwst\_miri\_apcorr\_0005.fits} for MIRI.
		
		Background emission in the images is modeled using \texttt{starbug2 --background} to produce clean images suitable for further analysis. Subsequently, PSF photometry is applied to these background-subtracted images using \texttt{starbug2 --psf}, in conjunction with WEBBPSF \citep{Perrin2014} version 1.1.1. Instrumental zero points are estimated using \texttt{starbug2 --calc-instr-zp}, which involves comparing PSF-based magnitudes with those obtained from reliable aperture photometry. The PSF photometry results are initially calibrated to the AB magnitude system using these zero points. These are then converted to the Vega magnitude system using the offsets provided in \texttt{jwst\_nircam\_abvegaoffset\_0001.asdf} for NIRCam and \texttt{jwst\_miri\_abvegaoffset\_0001.asdf} for MIRI. Subsequently, individual catalogs are merged using the routine \texttt{starbug2-match}. 
		
		\section{Results and Discussion} \label{sec:results}
		
\begin{figure*}
			\centering
			\includegraphics[scale=0.49]{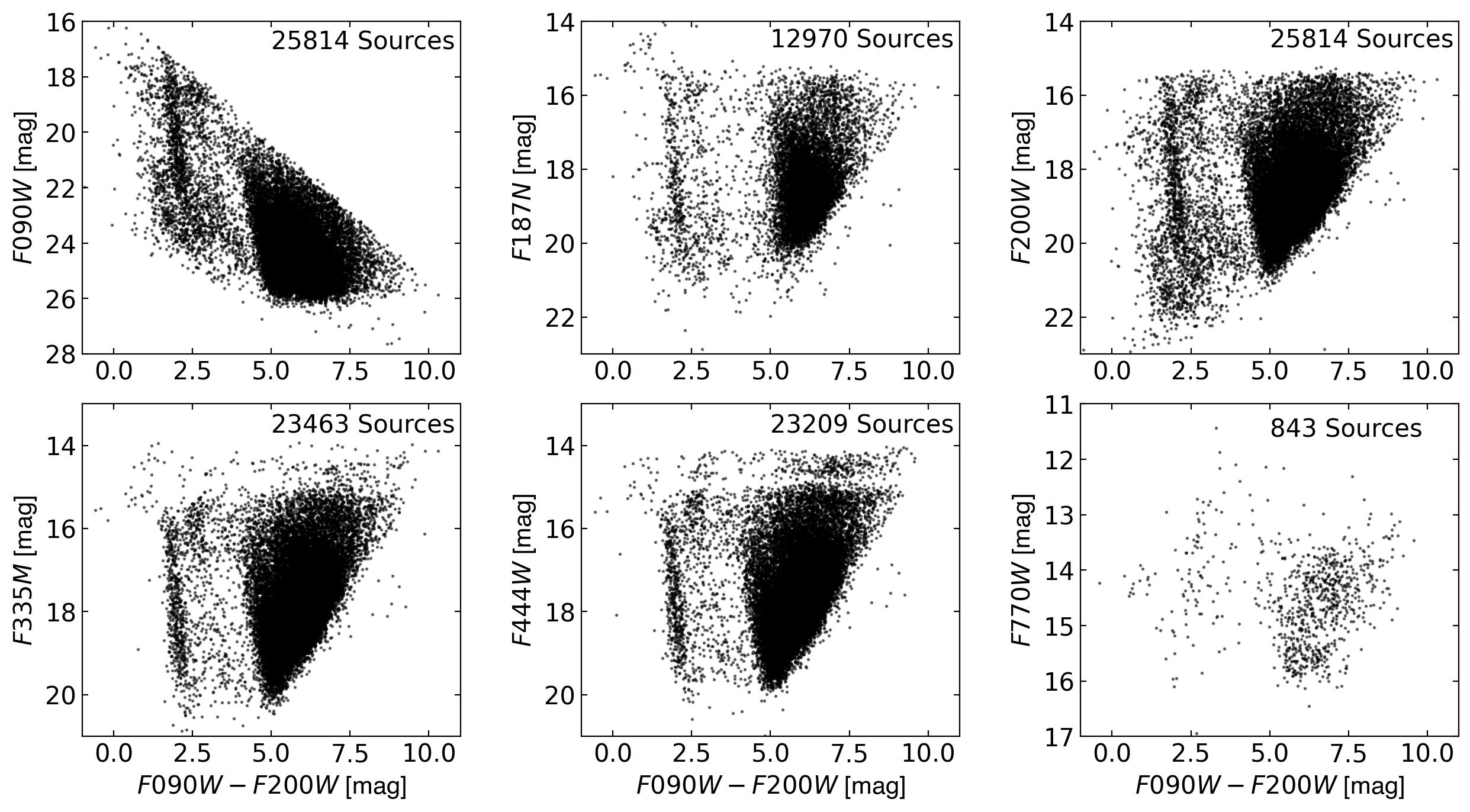}
			\caption{Color-magnitude diagrams for all identified sources within the Pillars of Creation in M16, depicted as black points. Each panel displays data obtained with a different filter $\lambda$ plotted against the color index $F090W-F200W$, where $\lambda$ includes $F090W$, $F187N$, $F200W$, $F335M$, $F444W$, and $F770W$. The total number of sources in each filter is indicated in the upper right corner of each panel.
				\label{fig:color_mag_diagram}}
		\end{figure*}

		Fig.~\ref{fig:color_mag_diagram} presents color-magnitude diagrams (CMDs) for the JWST NIRCam and MIRI data, plotting various filters ($\lambda$) against the color index $F090W-F200W$. These filters include $F090W$, $F187N$, $F200W$, $F335M$, $F444W$, and $F770W$. The CMDs are based on the final merged photometric catalog. Photometric uncertainties are typically below 0.2\,mag for NIRCam bands, while MIRI bands exhibit higher uncertainties due to their relatively lower signal-to-noise ratio (S/N). The diagrams distinctly show two prominent groups of stars, characterized by their varying color indices. The first group, comprising highly reddened young stars embedded within molecular clouds, exhibits a color index $F090W-F200W$ exceeding 5\,mag. The second group consists of foreground stars, with a color index approximately 2.5\,mag. Notably, the diagrams indicate a larger population of sources detected in the NIRCam bands compared to the MIRI bands.%, reflecting the NIRCam's broader sensitivity and lower noise characteristics.
		
		\subsection{Analysis of Color-Color Diagrams and Reddening Slopes}
		
		In the NIR and MIR bands, color-color diagrams are pivotal in understanding the extinction law. In these diagrams, stars mainly share similar intrinsic colors and they are often aligned along a path defined by the reddening vector which quantifies the displacement due to dust extinction. In areas with heavy dust obscuration, the natural color variations among stars are minor relative to the color excess from dust. This significant difference means that the variability in intrinsic colors can generally be overlooked, streamlining the analysis. This method is especially useful in regions of high extinction, where the effects of reddening overshadow any other changes in stellar colors \citep{Roman2007,Ascenso2013}. When evaluating two color indices, $m_{\lambda_1}-m_{\lambda_2}$ and $m_{\lambda_2}-m_{\lambda_3}$, the ratio of color excesses, $\frac{E(m_{\lambda_1}-m_{\lambda_2})}{E(m_{\lambda_2}-m_{\lambda_3})}$, can be derived through a linear regression of the stellar distribution on the diagram. The color excess, denoted as $E(m_{\lambda_1} - m_{\lambda_2})$, represents the deviation of the observed color from its intrinsic value, and is defined by: $E(m_{\lambda_1} - m_{\lambda_2}) = (m_{\lambda_1}-m_{\lambda_2})_{\text{observed}} - (m_{\lambda_1}-m_{\lambda_2})_{\text{intrinsic}}$.

		The effectiveness of these diagrams in delineating reddening slopes is influenced by the observational constraints across different filters. The observations at longer wavelengths can detect the stars with larger dust column density thanks to the much smaller extinction at the longer wavelengths. This results in variations in the depth of extinction that can be probed in different colors, as well as the sensitivity of these colors to extinction effects. In the subsequent sections, we will explore multi-band color-color diagrams to assess the consistency and variability of color excess ratios across different wavelengths, thereby enhancing our understanding of the interstellar dust in various environments.

		\subsubsection{$F090W-\lambda$ vs. $F090W-F200W$}
		
\begin{figure*}
			\centering
			\includegraphics[scale=0.24]{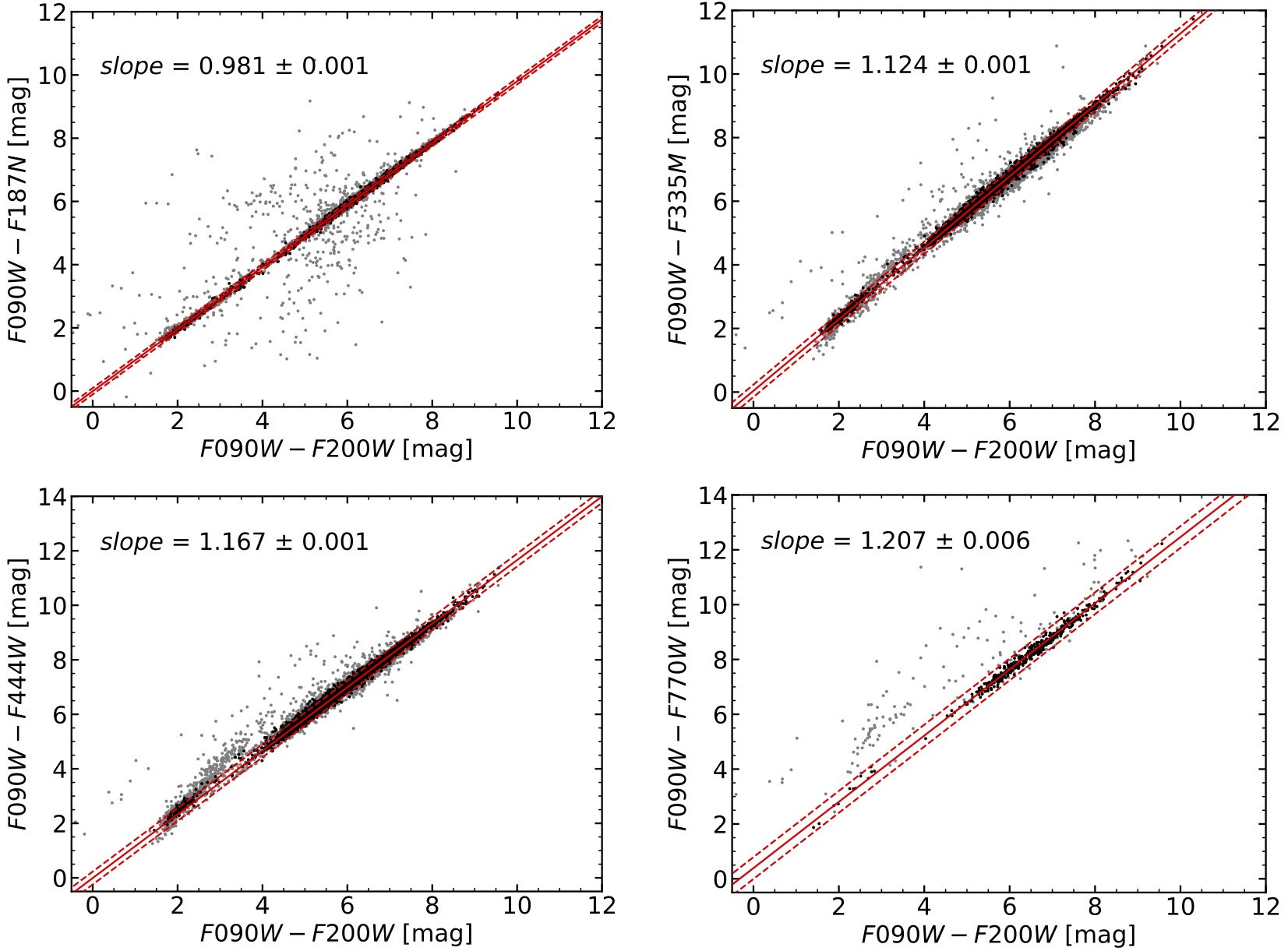}
			\caption{Color-color diagrams of $F090W-\lambda$ versus $F090W-F200W$ for all sources within the Pillars of Creation in M16. Panels display $\lambda$=$F187N$ (\emph{top left}),  $F335M$ (\emph{top right}), $F444W$ (\emph{lower left}), and $F770W$ (\emph{lower right}). The red solid lines represent the best-fit linear models to the color indices for each band using \texttt{LTS\_LINEFIT}. The $3\sigma$ confidence intervals of these fits are indicated by red dashed lines. Outliers are marked as grey points. Best-fit slopes are annotated in the upper left corner of each panel.
				\label{fig:color_color_diagram} }
		\end{figure*}

		We first select $F090W-F200W$ as the reference color to construct color-color diagrams for $\lambda = F187N, ~F335M, ~F444W,$ and $F770W$ (Fig.~\ref{fig:color_color_diagram}). Photometric uncertainties are constrained to less than 0.5\,mag for each band. The MIRI bands $F1130W$ and $F1500W$ are excluded due to the limited number of stellar detections, which precluded a reliable determination of color excess ratios.
		
		A linear regression is applied to the color-color diagrams using the \texttt{LTS\_LINEFIT} method, which accommodates errors in both color axes and intrinsic scatter within the diagram. This robust fitting technique efficiently identifies and excludes significant outliers (gray points in Fig.~\ref{fig:color_color_diagram}), focusing the fit on the retained black points. The computed color excess ratios, $\beta_\lambda = E(F090W-\lambda)/E(F090W-F200W)$, are summarized in Table~\ref{tab:ccr_slopes}.
		
		\begin{table*}
			\centering
			\caption{Derived color excess ratios $\beta_\lambda = E(F090W-\lambda)/E(F090W-F200W)$ across various bands for the Pillars of Creation in M16.}
			\label{tab:ccr_slopes}
			\begin{tabular}{c c c c c c}
				\hline \hline
				Band & $\lambda_{\rm eff}~(\mu m)$ & $1/\lambda_{\rm eff}~(1/\mu m)$ & This Work & \citet{Fahrion2023} & \citet{Wang2024} \\
				\hline
				$F090W$  & 0.899 &1.112 & 0 & 0  & 0  \\
				$F187N$  &  1.874 & 0.534 & 0.981$\pm$0.001 &0.954$\pm$0.162 & 0.968$\pm$0.002\\
				$F200W$ &  1.968 & 0.508 & 1  &  1 &1  \\
				$F335M$  &  3.354 & 0.298 &  1.124$\pm$0.001 &1.223$\pm$0.159 & 1.172$\pm$0.002\\
				$F444W$ & 4.350 & 0.230 & 1.167$\pm$0.001& 1.275$\pm$0.157 &1.213$\pm$0.002\\
				$F770W$ &  7.522& 0.133 &  1.207$\pm$0.006 &-- & --  \\
				\hline
			\end{tabular}
		\end{table*}
		
		Table~\ref{tab:ccr_slopes} also presents recent findings on extinction laws derived from observations by the JWST. \citet{Wang2024} utilized NIRSpec spectroscopic data to explore the NIR extinction law in the young massive star cluster Westerlund~2. Concurrently, \citet{Fahrion2023} analyzed NIRCam photometric data to probe the extinction law within the star-forming region 30 Doradus in the Large Magellanic Cloud (LMC). Our study reports color excess ratios $\beta_{F335M}$ and $\beta_{F444W}$ that are lower than those found in these studies, indicative of a comparatively flatter extinction law. 
		
\begin{figure*}
			\centering
			\includegraphics[scale=0.55]{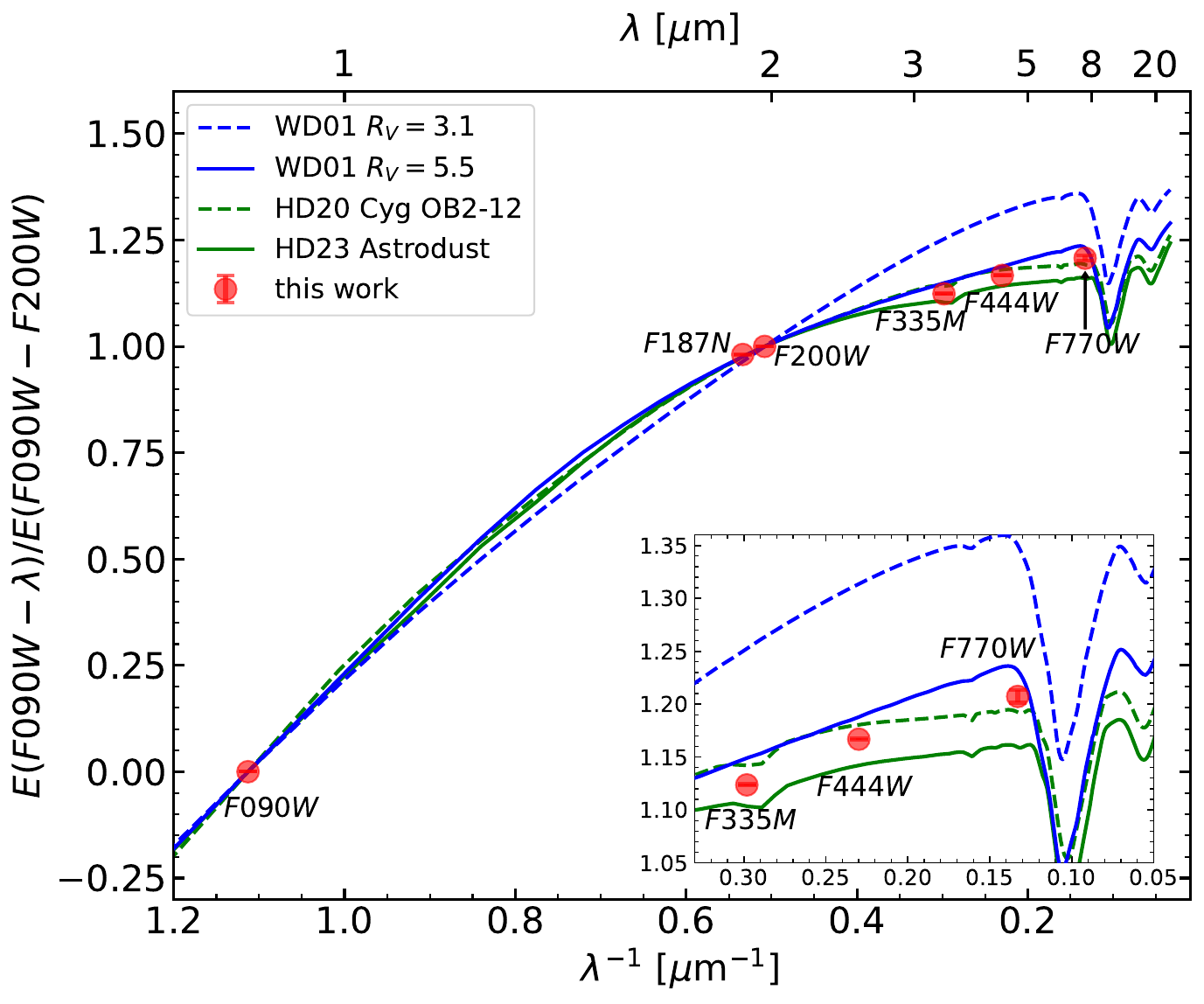}
			\caption{Comparison of the reddening law in M16 with literature results. Our derived color excess ratios (red points) are plotted alongside results from \citet[][WD01]{Weingartner2001} for $R_V=3.1$ in blue dashed and $R_V=5.5$ in blue solid, and from \citet[][HD20 Cyg OB2-12]{Hensley2020} in green dashed and \citet[][HD23 Astrodust]{Hensley2023} in green solid, respectively. The inset is a zoom-in view of the MIR regime.}
			\label{fig:reddening_law}
		\end{figure*}

		Fig.~\ref{fig:reddening_law} illustrates the derived color excess ratios, $\beta_\lambda$, plotted as a function of wavelength. Unlike previous studies that convert these ratios into relative extinction ratios, such as $A_\lambda/A_K$, which depend on uncertain NIR extinction ratios like $A_H/A_K$, we directly present $\beta_\lambda$. This approach minimizes errors associated with uncertainties in $A_H/A_K$. For comparison, Fig.~\ref{fig:reddening_law} includes reddening laws from from some important literatures. Our observed $\beta_\lambda$ curve is significantly flatter than the WD01 model for a diffuse environment at $R_V=3.1$, and even flatter than the WD01 model at $R_V=5.5$ for a dense environment, indicating a distinct extinction law in M16. Our results are more consistent with the extinction curve in diffuse ISM toward Cyg OB2-12 reported by \citet{Hensley2020} and the recent ``Astrodust" model by \citet{Hensley2023}. Specifically, the observed $\beta_\lambda$ is only slightly flatter than that reported by \citet{Hensley2020}, and slightly steeper than that of \citet{Hensley2023}. However, it is important to note that Cyg OB2-12 represents only a single case along a sightline through diffuse ISM, while M16 is a case for a dense region. More cases are required to confirm these results, and currently, we are conducting a systematic analysis of a sample of dense dark clouds to further substantiate our findings.

		The flatter extinction curves observed in our study suggest variations in dust grain size and composition within the dense regions of M16. WD01 suggest that a flatter infrared reddening curve correlates with larger grain size distributions. Additionally, the formation of ice mantles on dust grains, as discussed by \cite{Ormel2011} and \cite{McClure2023}, could further influence the extinction law characteristics. Recent observations by \cite{Ginsburg2023} demonstrated how ice absorption significantly alters star distributions on color-color diagrams in environments like the Galactic Center, as observed with JWST narrow-band photometry. These phenomena can lead to anomalous extinction effects in broadband filters, as noted in studies by \cite{Wang2013} and \cite{Xue2016}. In our data, the $F335M$ filter marginally covers the H$_2$O 3.0\,$\mu$m absorption feature, and the $F444W$ band seems unaffected by the CO$_2$ 4.27\,$\mu$m absorption, indicating that other factors may contribute to the observed flat extinction curve. This flat curve implies either a predominance of larger dust grains or a higher proportion of such grains in the dense M16 environment. This evidence reinforces the hypothesis that dust grain characteristics significantly influence the observed NIR extinction laws in various interstellar environments.

		\subsubsection{$F200W-F335M$ vs. $F335M-F444W$ and $F090W-F200W$}
		
\begin{figure*}
			\centering
			\includegraphics[scale=0.6]{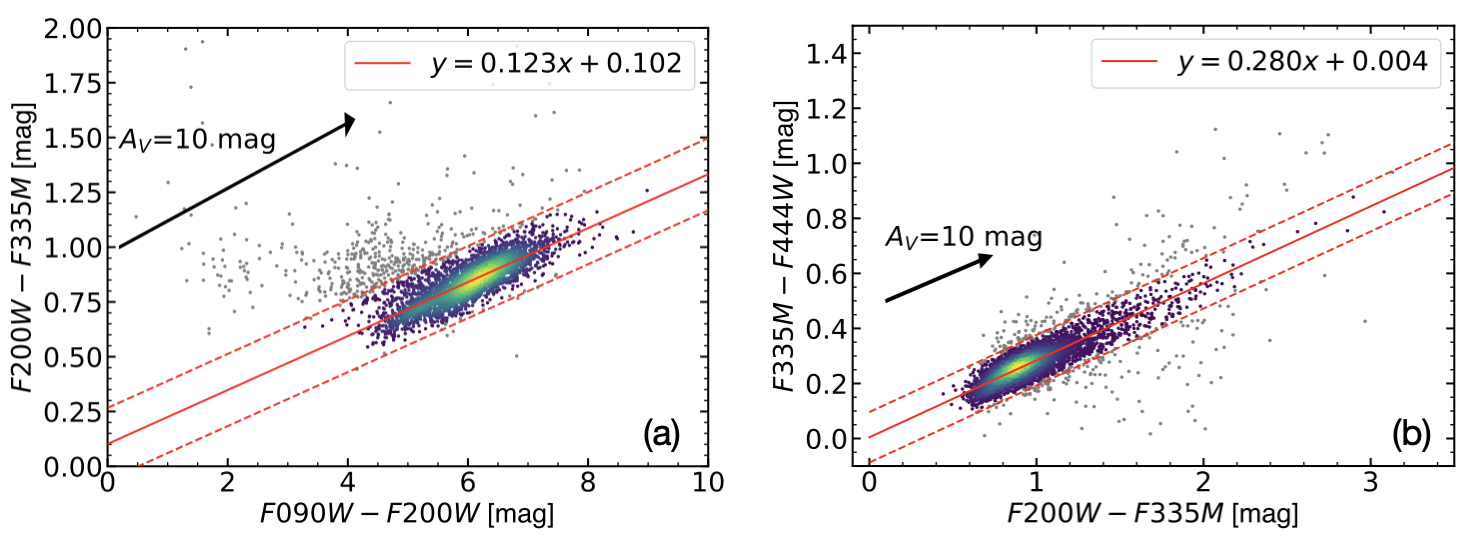}
			\caption{Color-color diagrams of $F200W-F335M$ vs. $F090W-F200W$ (panel a) and $F335M-F444W$ vs. $F200W-F335M$ (panel b), generated using color maps. The linear fits are shown with solid red lines, and the $3\sigma$ deviation ranges with dashed red lines. Outliers identified by \texttt{LTS\_LINEFIT} are marked with grey points. Reddening vectors for $A_V=10$\,mag, calculated using the WD01 $R_V=5.5$ extinction law, are superimposed.
				\label{fig:ccd_f200w_f335m} }
		\end{figure*}

		The M16 cloud exhibits significant dust obscuration, rendering the $F090W$ band largely ineffective for penetrating the dense regions. By contrast, the color indices $F200W-F335M$ and $F335M-F444W$ demonstrate enhanced capability in detecting the more embedded background stars, thanks to their reduced sensitivity to extinction compared to $F090W-F200W$. On the other hand, this leads to broader dispersions in color-color diagrams when using the $F200W-F335M$ and $F335M-F444W$ colors. To mitigate this dispersion, we have adopted the method outlined by \citet{Cambresy2011}, setting a pixel size of 4$^{\prime\prime}$ and applying a 3$\sigma$ clipping algorithm to compute the average color within each pixel. Subsequently, we have generated color maps for $F090W-F200W$, $F200W-F335M$, and $F335M-F444W$ colors. The resulting color-color diagrams, contrasting $F200W-F335M$ vs. $F090W-F200W$ and $F335M-F444W$ vs. $F200W-F335M$, are shown in Fig.~\ref{fig:ccd_f200w_f335m}. These diagrams are constructed using colors averaged from the maps, rather than relying on individual stellar observations. 
		
		As depicted in Fig.~\ref{fig:ccd_f200w_f335m}(a) and Fig.~\ref{fig:color_color_diagram}, the color index $F090W-F200W$ spans a range of 0 to 9\,mag. Considering a K0III star with an effective temperature of 4750\,K, its intrinsic color index $(F090W-F200W)_0$ is approximately 1.2\,mag \citep{Castelli2003}. Utilizing this intrinsic color and applying the WD01 $R_V=5.5$ model, the extinction depth $A_V$ for an embedded background star is calculated as $\sim23$\,mag. This is substantial yet typical for such dense regions. A linear regression of $F200W-F335M$ vs. $F090W-F335M$ indicates a color excess ratio $E(F200W-F335M)/E(F090W-F200W) \approx 0.123$, closely matching the theoretical value of $\beta_{F335M}-\beta_{F200W} = 0.124$ derived earlier in Section 3.1.1, affirming the reliability of our methodology.
		
		Selecting longer wavelength bands such as $F200W-F335M$ over $F090W-F200W$ enables us to discern the extinction law within the densest regions of M16. Fig.~\ref{fig:ccd_f200w_f335m}(b) shows that $F200W-F335M$ spans a range of 0--3\,mag. Using the WD01 model with $R_V = 5.5$ for a K0III giant star, the maximum depth of extinction reaches approximately $A_V \sim 60$ mag. The observed color excess ratio curve, as depicted in Fig.~\ref{fig:reddening_law}, is flatter than that predicted by this model, suggesting that the actual depth could be even higher. At this depth, the derived color excess ratio is $E(F335M-F444W)/E(F200W-F335M) \approx 0.28$. In contrast, using $F090W-F200W$ as the reference, the color excess ratio is computed as $\left(\beta_{F444W} - \beta_{F335M}\right) / \left(\beta_{F335M} - 1\right) \approx 0.347$, which is higher than the 0.28 obtained from direct linear fitting of $F335M-F444W$ versus $F200W-F335M$. This systematic difference suggests that discrepancies may arise due to changes in dust properties with increased extinction depth, potentially indicating larger dust grain sizes. Additional independent observational evidence for the growth of dust grains in dense interstellar environments comes from the observations of near-IR core shine \citep{Jones2016}, and of strongly forward-directed scattering phase functions, combined with a high long-wavelength albedo of dust in globules \citep{Witt1990,Togi2017}.

		\subsection{Variability of the power-law index $\alpha$ in the NIR extinction Law}
		
		Numerous studies have demonstrated that the NIR extinction law can be characterized as a power-law form $A_\lambda\propto\lambda^{-\alpha}$. Pioneering work by \citet{Naoi2006,Naoi2007} highlighted that the NIR extinction slope tends to flatten with increasing extinction depth. Further, \citet{Nogueras2019} documented variations in $\alpha$ between the NIR $JH$ and $HK$ bands, with $\alpha_{JH}$ consistently exceeding $\alpha_{HK}$. These observations suggest a decrease in the extinction law slope at longer wavelengths. Recent spectroscopic studies extend these findings across the 0.8 to 5.0\,$\mu$m range, confirming the power-law behavior with varying indices \citep{Decleir2022,Wang2024}.
		
		Despite the limitations of photometric data, our analysis leverages multi-wavelength color excess ratios to explore the $\alpha$ variability. Assuming a power-law model for NIR extinction, we derive the relationship between the color excess ratios and $\alpha$ as follows:
		\begin{equation} \label{equ:alpha}
			\frac{E(m_{\lambda_2}-m_{\lambda_3})}{E(m_{\lambda_1}-m_{\lambda_2})} = \frac{\lambda_2^{-\alpha} - \lambda_3^{-\alpha}}{\lambda_1^{-\alpha} - \lambda_2^{-\alpha}}.
		\end{equation}
		Here, $\lambda_1$, $\lambda_2$, and $\lambda_3$ denote the effective wavelengths of the JWST NIRCam bands, ordered as $\lambda_1 < \lambda_2 < \lambda_3$. Fig.~\ref{fig:ccd_f200w_f335m}(a) presents the fitting of the color-color diagram for $F200W-F335M$ versus $F090W-F200W$. From this, we derived a color excess ratio $E(F200W-F335M)/E(F090W-F200W) = 0.123$. Using Equation~\ref{equ:alpha}, we estimate that $\alpha_{F090W\sim F335M} = 2.44$.
		
		Additionally, as depicted in Fig.~\ref{fig:ccd_f200w_f335m}(b), the analysis of $F335M-F444W$ versus $F200W-F335M$ yielded a color excess ratio $E(F335M-F444W)/E(F200W-F335M) = 0.28$. This corresponds to $\alpha_{F200W\sim F444W} = 1.63$, markedly lower than $\alpha_{F090W\sim F335M} = 2.44$. These findings affirm the non-uniformity of $\alpha$ across the spectrum from 0.9\,$\mu$m to 4.4\,$\mu$m, underscoring a trend toward flatter extinction curves at longer wavelengths. Moreover, in the MIR wavelengths, compared to the NIR, we are able to probe denser regions of molecular clouds that exhibit higher extinction values. The observed lower values of $\alpha$ in the MIR, indicative of flatter extinction curves relative to those in the NIR, further support our previous observations that denser regions within star-forming environments tend to exhibit flatter extinction curves.
		
		\section{Summary} \label{sec:summary}
		
		This study utilizes the sophisticated imaging capabilities of the JWST NIRCam and MIRI to conduct deep observations of the Pillars of Creation within the M16 star-forming region. Our analysis has yielded the color-excess ratio curve across a wavelength range of 0.9 to 8\,$\mu$m. By conducting linear fits on the color-color diagrams, we are able to directly derive the extinction law in M16. The enhanced detection capabilities of JWST data allow us to probe extinction depths up to $A_V \approx 60$\,mag. The derived color-excess ratio curve for the dense region in M16 exhibits a flatter profile than those typically obtained from previous works, and is even less steep than the WD01 $R_V=5.5$ extinction curve.
		
		Furthermore, our findings reveal that the extinction law in the MIR range from 2 to 5\,$\mu$m does not simply continue the NIR power-law relationship. Instead, we observed a decrease in the slope of the IR extinction law with increasing wavelength, suggesting a modification in dust characteristics or scattering processes in the MIR spectrum.
		
		\begin{acknowledgments}
			We would like to thank the anonymous referee for the very helpful comments and suggestions. This work is supported by the National Key R\&D program of China 2022YFA1603102 and 2019YFA0405500, the National Natural Science Foundation of China 12133002, 12173034, and 12322304. B.Q.C. acknowledges the National Natural Science Foundation of Yunnan Province 202301AV070002 and the Xingdian talent support program of Yunnan Province. X.C. thanks to Guangdong Province Universities and Colleges Pearl River Scholar Funded Scheme (2019).
		\end{acknowledgments}
		
		\vspace{5mm}
		\facilities{{\em JWST} (NIRCam \& MIRI) - James Webb Space Telescope.}
		
		\software{JHAT \citep{Rest2023},
			image1overf.py (\citep{Willott2022}, {\sc starbugii} \citep{Nally2023} , LtsFit \citep{Cappellari2014}.
		}
		
		\bibliography{m16-ext}{}
		\bibliographystyle{aasjournal}
		
	\end{CJK*}
\end{document}